# A Deep Learning Approach to Automate High-Resolution Blood Vessel Reconstruction on Computerized Tomography Images With or Without the Use of Contrast Agent

Anirudh Chandrashekar[1#], Ashok Handa[1#],
Natesh Shivakumar[1], Pierfrancesco Lapolla[1],
Vicente Grau[2*], Regent Lee[1*]

1. Nuffield Department of Surgical Sciences, University of Oxford
2. Department of Engineering Science, University of Oxford

#*Authors contributed equally

Corresponding Author:
Regent Lee
Regent.Lee@nds.ox.ac.uk

## Abstract

Existing methods to reconstruct vascular structures from a computed tomography (CT) angiogram rely on injection of intravenous contrast to enhance the radio-density within the vessel lumen. However, pathological changes can be present in the blood lumen, vessel wall or a combination of both that prevent accurate reconstruction. In the example of aortic aneurysmal disease, a blood clot or thrombus adherent to the aortic wall within the expanding aneurysmal sac is present in 70-80% of cases. These deformations prevent the automatic extraction of vital clinically relevant information by current methods. In this study, we implemented a modified U-Net architecture with attention-gating to establish a high-throughput and automated segmentation pipeline of pathological blood vessels in CT images acquired with or without the use of a contrast agent. Twenty-six patients with paired non-contrast and contrast-enhanced CT images within the ongoing Oxford Abdominal Aortic Aneurysm (OxAAA) study were randomly selected, manually annotated and used for model training and evaluation (13/13). Data augmentation methods were implemented to diversify the training data set in a ratio of 10:1. The performance of our Attention-based U-Net in extracting both the inner lumen and the outer wall of the aortic aneurysm from CT angiograms (CTA) was compared against a generic 3-D U-Net and displayed superior results. Subsequent implementation of this network architecture within the aortic segmentation pipeline from both contrast-enhanced CTA and non-contrast CT images has allowed for accurate and efficient extraction of the entire aortic volume. This extracted volume can be used to standardize current methods of aneurysmal disease management and sets the foundation for subsequent complex geometric and morphological analysis. Furthermore, the proposed pipeline can be extended to other vascular pathologies.

**Introduction**

A computerised tomography (CT) scan uses computer-processed combinations of multiple X-ray measurements taken from different angles to produce cross-sectional images (virtual "slices") of specific areas of a scanned object. This allows visualisation inside the object without cutting it open. Since the invention of the first commercially available CT scanner in 1972[1], the use of CT scans for the diagnosis and management of disease is extensively embedded in every field of modern medicine. In the NHS alone, ~6 million CT scans were performed in 2018-2019[2].

Visualisation of blood vessels on a routine CT scan is challenging. Blood vessels consist of vessel wall structures, and the contents within the vessel lumen (blood, clot, plaques, etc). These components have similar radio-densities (measured in Hounsfield Unit, HU) to the adjacent soft tissue structures. There are no existing automated methods to segment blood vessels on CT scans without the use of an intravenous contrast agent. Injection of intravenous contrast enhances the radio-density within vessel lumens and enables its reconstruction. The produced CT angiogram is routinely utilised to diagnose medical problems related to blood vessels.

However, pathological changes can be present in the blood lumen, vessel wall or a combination of both. In the example of aortic aneurysms (AA, abnormal ballooning of the main artery 'aorta' in the body) (Figure 1A, red arrow), there is usually a blood clot or thrombus adherent to the aortic wall within the aneurysm sac (Figure 1B, red arrow points toward the AA). Existing automated methods to reconstruct the angiogram would isolate the inner lumen but are unable to extract the thrombus and the complex thrombus-lumen interface. As such, there is no automated method to assess the aneurysm diameter (Figure 1C) or thrombus volume. These are vital information used in the clinical care and research of patients with abdominal aortic aneurysms.

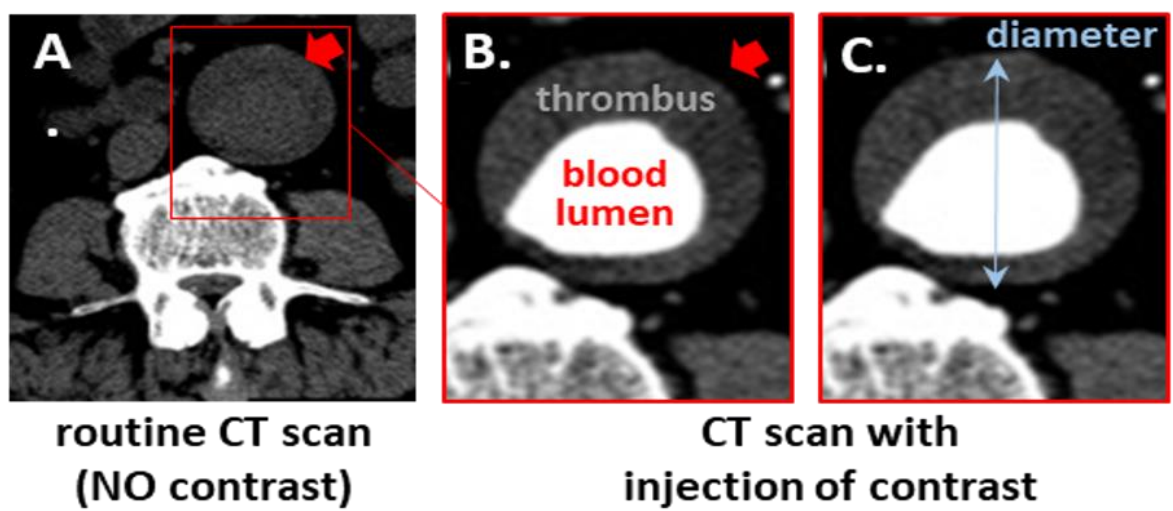

**Figure 1:** Axial slice through an abdominal aortic aneurysm from a non-contrast CT and CTA (+ contrast injection) image.

Prior to the advent of deep learning, CT image segmentation methods for vascular imaging were built on traditional tools including edge detection and/or mathematical models. These methods are complex, difficult to execute and are often poorly generalizable. Subsequent machine learning methods rely on a pattern recognition system, which involves feature extraction. This process, known as "hand-crafting" features, involves devising an algorithm to extract a feature representation of the given data. This representation, which is difficult to compile, then serves as the criteria to achieve the given task[3].

In the early 2000s, deep learning (DL) methods became more approachable for image processing tasks, given significant improvements in hardware. Convolutional neural networks (CNNs) are the foundation of many DL architectures and consists of multiple layers involved in transforming the input using various pre-defined methods (convolution, non-linear activation, pooling, loss calculation, etc). High-level abstractions attained by this process are then extracted by fully connected layers. Finally, the weights of each neural connection and by extension the model are optimized during model training[3, 4]. In recent years, many groups working in this domain have strived to identify improvements to this conventional approach for their segmentation task.

One of the most well-known architectures that uses many of these principles for biomedical image segmentation tasks is the U-Net[5]. A major advantage of this model is the use of skip connections between different stages of the network. This serves to integrate the spatial and contextual information to assemble a more precise output. Furthermore, these methods, which were initially limited to 2D slices, have since been applied to 3D image volumes to fully utilize the advantage of spatial information[3, 5]. Several limitations of current 3D U-Net methods have restricted its implementation for CT images. Due to memory limitations for model training, current 3D U-Net methods utilise down-sampled input images. Therefore, this input size may not have enough resolution to represent the diverse anatomical image in a singular image. This is especially relevant when assessing structures with variation that can only be captured at higher resolution[3, 6]. Additionally, most methods are not automatic and are unable to produce high-resolution segmentations from the given high-resolution input.

In this study, we implement a modified U-Net architecture to achieve high-throughput, automated segmentation of blood vessels in CT images acquired with or without the use of contrast agent. In contrast enhanced CT images, our U-Net architecture further enables simultaneous segmentation of both the arterial wall and blood flow lumen to enable characterisation of the pathological contents. We demonstrate the efficacy of this U-Net architecture by reconstructing the thoracic and abdominal aorta, which is the main artery bringing blood supply from the heart to the rest of the body.

**Methods**
**CT images from a clinical cohort**
Computerized Tomographic scans of the chest and abdomen were acquired through the Oxford Abdominal Aortic Aneurysm (OxAAA) study. The study received full regulatory and ethics approval from both Oxford University and Oxford University Hospitals (OUH) National Health Services (NHS) Foundation Trust (Ethics Ref 13/SC/0250). As part of the routine pre-operative assessment for aortic aneurysmal disease, a non-contrast CT of the abdomen and a CT angiogram (CTA) of both the chest and abdomen was performed for each patient. CTA images were obtained following contrast injection in helical mode with a pre-defined slice thickness of 1.25 mm. Non-contrast CT images included only the descending and abdominal aorta and were obtained with a pre-defined slice thickness of 2.5 mm.

Paired contrast and non-contrast CT images were anonymised within the OUH PACS system before being downloaded onto the secure study drive.

**Manual Segmentation of CT Images**

Twenty-six patients with paired non-contrast and CTA images of the abdominal region were randomly selected. In the CTA, both the aortic inner lumen and outer wall were segmented from the aortic root to the iliac bifurcation using the ITK-Snap segmentation software[7].

Semi-automatic segmentation of the aortic inner lumen was achieved using a variation of region-growing by manually delimiting the target intensities between the contrast-enhanced lumen and surrounding tissue. Segmentation of the aortic outer wall was performed manually by drawing along its boundary using the previously obtained inner lumen as a base. Removing the inner lumen from the larger outer wall segmentation results in a segmentation mask highlighting the content between the arterial wall and blood lumen (in this case, thrombus). In the non-contrast CT image, the aorta was manually segmented.

The left most panel in Figure 2 depict axial slices from the CT of chest and abdomen. Figures 2A and 2B are CTA images depicting the ascending thoracic aorta (yellow arrow), descending thoracic aorta (blue arrow), and abdominal aorta (red arrow) which is aneurysmal and contains crescentic layers of thrombus. Figure 2C is the corresponding cross section of the abdominal aorta (red arrow) in the non-contrast CT scan. The aortic contour is manually segmented using IKT snap. Figure 2 D, E, F, show the CT images with the overlying manually derived segmentation. Views of the 3D volumes derived from the manual 2D segmentations are depicted in Fig 2 G-H.

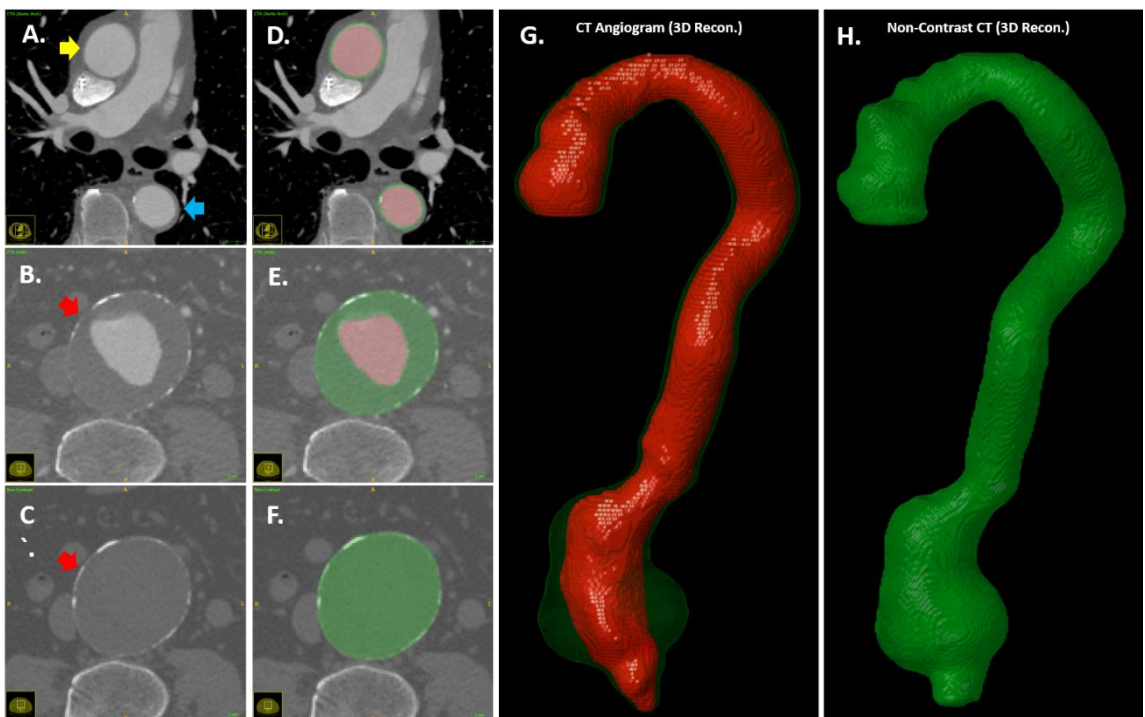

**Figure 2: A-F**. Axial Slices obtained from a CTA and non-contrast CT scan with overlaying manually segmented labels. **G-H.** Corresponding 3D-reconstructed volumes are generated from the image masks.

**Assessment of intra- and inter- observer variation of manual segmentation**

A subset of these scans was selected randomly for intra- and inter-observer variability evaluation (n = 10). This directly assessed the validity and accuracy of the manual segmentations used for subsequent analysis. For the intra-observer assessment, manual segmentation of the 10 scans was performed for the second time after a gap of 2 weeks. For the inter-observer assessment, a trained clinician performed the segmentations independent of the primary observer. In both instances, segmentation masks were compared against the ground truth (observer 1).

**Data augmentation**

Of the 26 patients, 13 patients were randomly allocated to the training ($n_{train}$ = 10) and validation cohorts ($n_{train}$ = 3). Following manual segmentation, the original CT images and their corresponding image masks of only patients in the training/validation cohorts were augmented using divergence transformations. In order to diversify the training data set, divergence transformations employ non-linear warping techniques to each axial slice, which manipulate the image in a certain predefined location. In figure 3A, the aneurysmal sac can be seen towards the base of the gaussian peak and is noticeably stretched. By selecting the shape of the 3D surface, different warping affects can be created. In order to create localized stretching (Fig 2B) a 2-D gaussian curve is used:

$$g(i,j) = exp - \frac{(i-ic)^2 + (j-jc)^2}{2\sigma^2}$$

Here ($I_c, J_c$) is the centre from which the image is locally stretched. The images were augmented in this manner with gaussians at 5 locations adjacent to the aorta (Fig 3A). Furthermore, this method was extended to achieve both congruent and divergent local transformations. Therefore, each patient's scan in the training cohort was augmented 10:1 to obtain a total of 143 post-augmented scans. Figure 3 illustrates a single axial slice augmented 10 times. During training, each 3D image was augmented further using random rotation (0-15°), translation and scaling (0.7 - 1.3).

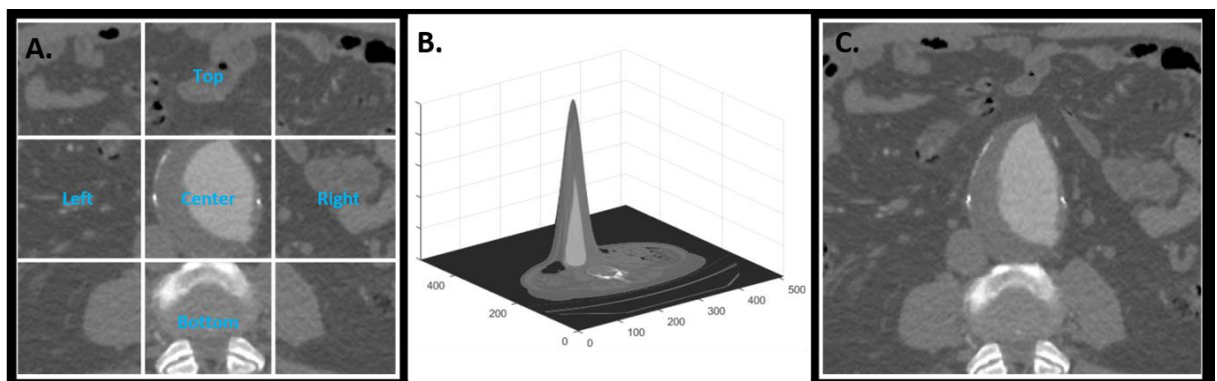

**Figure 3: A.** Axial slice from a contrast-enhanced CT scan with pre-defined locations for the divergence transformation function. **B.** The transformation employs non-linear warping techniques to create localized stretching. **C.** Augmented Axial Slice following divergence transformation.

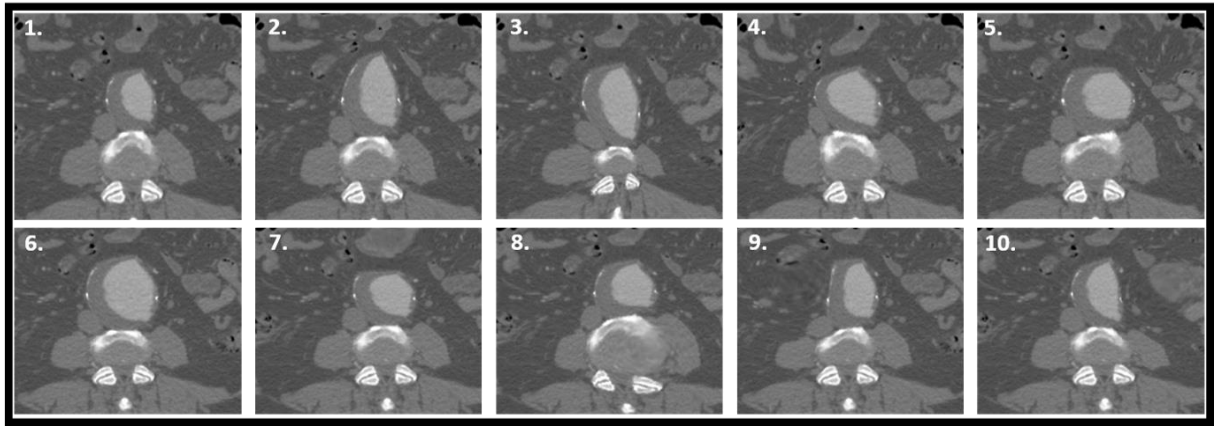

**Figure 4:** Sample axial slice is augmented 10:1 using divergence transformations.

As see in Table 1, post-augmented scans were split based on the original pre-augmented images into training ($n_{train}$ = 10 patients, 110 augmented scans), and validation ($n_{valid}$ = 3 patients, 33 augmented scans) groups. This was done to avoid data leakage or the intermingling of patients and their augmented scans in the training/validation groups. Here, the validation group was used at the end of each training epoch to gauge model performance and fine-tune model hyperparameters. The remaining 13 patients formed the testing cohort ($n_{test}$ = 13).

**Table 1:** Patient allocation between the training, validation and testing cohorts.

|  | **Model Training** | | **Model Evaluation** |
|---|---|---|---|
|  | Training Cohort ($n_{train}$) | Validation Cohort ($n_{valid}$) | Testing Cohort $n_{test}$ |
| **Patients** | 10 | 3 | 13 |
| **Post-Augmented Scans** | 110 | 33 | - |

**U-Net Architecture**

In this study, we utilised a variation of the U-Net for both the Aortic Region-of-interest (ROI) detection and segmentation tasks[5, 6]. The general architecture of the U-Net consists of two components: the contraction path and expansion path (Figure 5). The contraction path (red) serves to extract information and capture the context of the input at the expense of losing spatial information. Here, as the input CT image (CTA/Non-Contrast) is deconstructed, it is able to extract more complex features relevant to the aortic segmentation task. This is followed by an expansion path (green), where the size of the image gradually increases to produce a predictive binary mask. The lost spatial information is restored using skip connections and is merged via concatenation. These connect the output of the down-sampling path with the extracted feature maps/ input of the up-sampling path at the same level. This serves to integrate the spatial and contextual information to assemble a more precise prediction of the aortic structure.

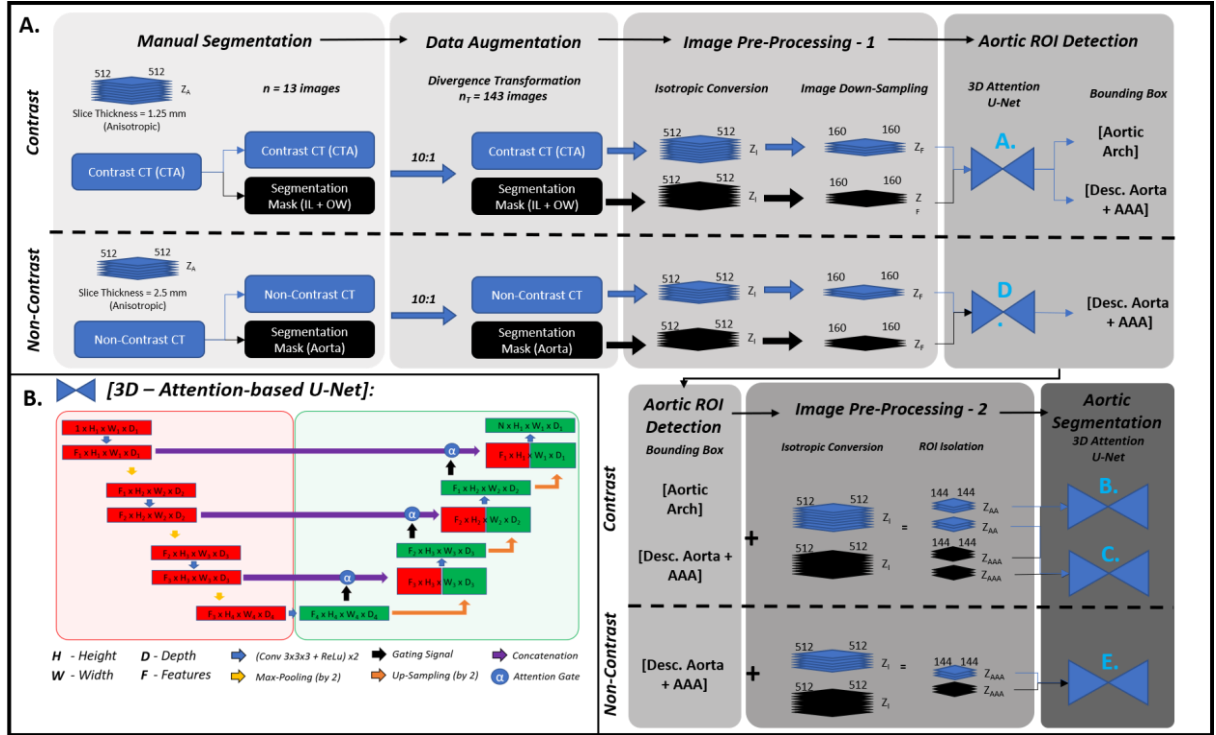

**Figure 5: A.** Automatic high-resolution Aortic segmentation pipeline for the simultaneous detection of the aortic inner lumen, and intra-luminal thrombus/outer wall. **B.** The base architecture for this pipeline is a 3-D Attention-based U-Net.

**Attention Gating**

We evaluated the use of a 3D U-Net with attention gating against a generic 3D U-Net for segmentation of the aorta. Information extracted from the coarse scale is used within this gating mechanism to filter out irrelevant and noisy data exchanged via the skip connections before the concatenation step. The output of each attention gate is the element-wise multiplication of input feature-maps and a learned attention coefficient [0 – 1]. Given that we are simultaneously predicting the location of the aortic inner lumen and outer wall, multi-dimensional attention coefficients were used to focus on a subset of target structures. The gating coefficients were determined using additive addition [8], which has been shown to be more accurate than multiplicative addition [9].

The integration of attention gates for the purpose of segmentation of pancreas has produced superior results when compared to that of prior models[6]. A similar attention gating mechanism was used in this study. In order to assess its benefit for AAA segmentation, the performance of an attention-based 3-D U-Net was compared against that of a generic 3-D U-Net.

Figure 5B illustrates the 3D U-Net architecture with the attention gates utilized in this study. The initial learning rate and weight decay for both models were set to $1.0 * 10^{-3}$ and $1.0 * 10^{-6}$, respectively. Training to segment the aneurysmal region was carried out for a total of 1000 epochs with a batch size of 2 3D Images.

**Loss Function**

To quantify the performance of our algorithm at each step, we utilized the DICE score, which is a well-known performance metric in image segmentation tasks. This metric gauges the similarity between two images (A and B) and is defined as follows:

$$Dice\ (A, B) = \frac{2|A \cap B|}{|A| + |B|}$$

**Aortic Segmentation Pipeline: Aortic ROI Detection**

Following data augmentation, all images were pre-processed. Pre-processing steps included isotropic voxel conversion and image down-sampling by a factor of 3.2 (512 x 512 x $Z_i$ → 160 x 160 x $Z_f$). This was performed to allow for increased efficiency during model training. The next step in this automatic aortic segmentation pipeline is Aortic ROI detection. This was performed on both the contrast and non-contrast CT images to isolate the aortic region for subsequent segmentation.

Attention U-Nets A and D *(Attn U-Net A, D)* were trained for a total of 600 epochs to segment the aorta from these decreased resolution, isotropic CTA and non-contrast CT images, respectively. The initial learning rate, weight decay, and batch-size for model training were set to $1.0 * 10^{-3}$ and $1.0 * 10^{-6}$ and 2 3D Images. Aortic bounding boxes were generated from the predicted segmentation masks. Two bounding boxes were generated from the contrast CT image (1. Aortic Arch and 2. Descending Aorta and AAA) and one was generated from the non-contrast CT image (1. Descending Aorta and AAA). Regions of interests (144 x 144 x [$Z_{AA}$ or $Z_{AAA}$]) centred around the defined bounding box were isolated and served as the input data for aortic segmentation.

**Aortic Segmentation Pipeline: Aortic Segmentation**

U-Nets B and C *(Attn U-Net B, C)* were trained for 1500 epochs on the CTA CT ROIs. They were tasked to simultaneously segment the aortic inner lumen and ILT/outer wall regions of the aortic arch and descending aorta/AAA, respectively. On the other hand, U-Net E was trained for 1000 epochs on the non-contrast ROIs and was tasked to segment the descending aorta/AAA. The learning rate, weight decay, and batch-size for all U-Nets were set to $1.0 * 10^{-3}$ and $1.0 * 10^{-6}$ and 2 3D Images, respectively.

Table 2 delineates all the U-Nets trained and evaluated in this study. Model training was performed simultaneously on a workstation with 2 11gb NVIDIA RTX 2080 TI graphics cards. Following training, in order to assess model performance and generalizability, models were evaluated on an external test cohort of non-augmented CTA and non-contrast images ($n_{ext}$= 13 scans). This cohort of scans was obtained from the same patient population and was independent of the scans used during training.

**Table 2**: All 3D U-Nets trained/evaluated in this study as depicted in **Figure 4**.

| Model | Epochs | Goal |
|---|---|---|
| *U-Net* | 1000 | Multi-Class AAA Segmentation |
| *Attn U-Net* | 1000 | Multi-Class AAA Segmentation |
| *Attn U-Net A* | 600 | Aortic Segmentation from low-resolution isotropic CTA |
| *Attn U-Net B* | 1500 | Multi-Class Aortic Arch Segmentation from high-resolution isotropic CTA |
| *Attn U-Net C* | 1500 | Multi-Class Descending Aorta + AAA Segmentation from high-res. isotropic CTA |
| *Attn U-Net D* | 600 | Aortic Segmentation from low-resolution isotropic Non-Contrast CT |
| *Attn U-Net E* | 600 | Aortic Segmentation from high-resolution isotropic Non-Contrast CT |

## Results

### CT Image Characteristics

Of the cases (n=26) included in the study, 13 were used for model training and the remaining 13 used for model testing. Details regarding the CT image characteristics between these groups are summarised in Table 3.

**Table 3**: Image characteristics within the training and external test cohorts.

| | | Training Cohort (n = 13) | | Test Cohort (n = 13) | | p-value |
|---|---|---|---|---|---|---|
| **Contrast** | 25th Percentile HU [95% CI] | -1008 | [-1003 -1014] | -1006 | [-1002 -1010] | 0.42 |
| | Mean HU [95% CI] | -587 | [-646.1 -527.6] | -568.4 | [-610.2 -526.6] | 0.48 |
| | 75th Percentile [95% CI] | -67 | [-45.8 -85.8] | -54.1 | [-40.8 -67.4] | 0.18 |
| | Standard Deviation [95% CI] | 484.1 | [475.0 493.2] | 490.6 | [485.1 495.9] | 0.08 |
| | Voxel Length [95% CI] | 0.81 mm | [0.76 0.86] | 0.83 mm | [0.79 0.87] | 0.50 |
| | Voxel Height [95% CI] | 0.81 mm | [0.76 0.86] | 0.83 mm | [0.79 0.87] | 0.50 |
| | Voxel Thickness | 1.25 mm | | 1.25 mm | | - |
| | KVP | 120 | | 120 | | - |
| | Exposure Time [95% CI] | 434 | [359.2 508.8] | 474.2 | [369.3 579.0] | 0.50 |
| **Non-Contrast** | 25th Percentile HU [95% CI] | -1009 | [-1005 -1013] | -1005 | [-1002 -1008] | 0.07 |
| | Mean HU [95% CI] | -565 | [-524.3 -606.7] | -550.6 | [-589.4 -511.9] | 0.55 |
| | 75th Percentile [95% CI] | -53.4 | [-39.9 -66.9] | -46.8 | [-62.2 -31.3] | 0.49 |
| | Standard Deviation [95% CI] | 483.2 | [476.0 490.0] | 484.8 | [479.8 489.9] | 0.69 |
| | Voxel Length [95% CI] | 0.80 mm | [0.75 0.85] | 0.82 mm | [0.78 0.86] | 0.58 |
| | Voxel Height [95% CI] | 0.80 mm | [0.75 0.85] | 0.82 mm | [0.78 0.86] | 0.58 |
| | Voxel Thickness | 2.5 mm | | 2.5 mm | | - |
| | KVP | 120 | | 120 | | - |
| | Exposure Time [95% CI] | 457.3 | [435.2 467.8] | 462.2 | [448.3 475.0] | 0.63 |

### *Intra-* and *Inter-* observer variability assessment

There were strong agreements for both *inter-* and *intra-* observer measurements (intra-class correlation coefficient, 'ICC' = 0.995 and 1.00, respective. P<0.001 for both). Table 4 summarises the DICE score metrics for the *intra-* and *inter-* observer variability assessments performed on CTA and non-contrast CT images. The inter-operator variability is greater than intra-operator variability for all regions, as seen by the lower DICE scores. These data supports the accuracy of the manual segmentations used for model training.

**Table 4**: DICE score metric for *Intra-*/*Inter-* operator aortic segmentations

| | Region | *Intra-* | *Inter-* |
|---|---|---|---|
| **Contrast** | Inner Lumen | 98.0 ± 0.2 % | 96.5 ± 0.4 % |
| | Entire Aorta | 97.8 ± 0.5 % | 96.1 ± 0.6 % |
| | Outer Wall + ILT Only | 95.1 ± 0.8 % | 93.1 ± 0.9 % |
| **Non-Contrast** | Entire Aorta | 96.8 ± 0.4% | 95.2 ± 0.8 % |

**Attention-based 3D-U-Net vs 3D-U-Net for AAA segmentation**

To assess the benefit of attention-gating for AAA segmentation, the performance of an attention-based 3-D U-Net was compared against that of a generic 3-D U-Net. Figure 6 illustrates the evolving DICE score metric for the validation group during model training. During the training of the Attention-based U-Net, the overall DICE score increased from 24% at epoch 1 (Inner Lumen: 36.7 %, Outer Wall: 11.9%) to approximately 95.3% at epoch 1000 (Inner Lumen: 97.4%, Outer Wall: 89.2%). On the other hand, the performance of the control 3D U-Net increased from 23.0% at epoch 1 (Inner Lumen: 38.2%, Outer Wall: 7.7%) to approximately 91.8.0% at epoch 1000 (Inner Lumen: 96.4%, Outer Wall: 87.2%).

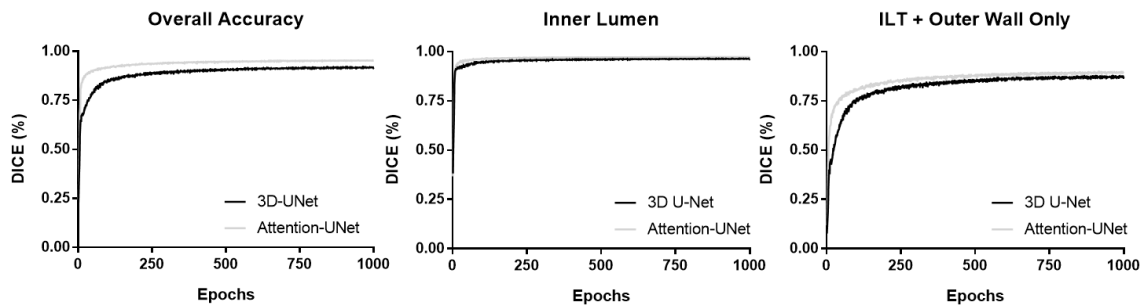

**Figure 6:** *Attention U-Net vs 3D U-Net for AAA Segmentation*. Training Paradigm for the Validation cohort. The Attention U-Net and generic 3D U-Net were trained for a total of 1000 epochs. Model outputs were assessed at each iteration and were compared against the ground truth segmentation using the DICE metric.

Segmentation of the testing cohort was used to evaluate model performance. Model output was compared against the manually segmented ground-truth images utilizing the DICE score metric. The results of this analysis are found in Table 3. The accuracy of the Attention-based U-Net in extracting both the inner lumen and the outer wall of the aneurysm is superior to that of the generic 3D- U Net. This rationalizes the incorporation of the attention-gating unit into the segmentation pipeline.

**Table 5:** Aortic Segmentation Accuracy (DICE Score)

| Region | *Attention U-Net* | *3D U-Net* |
|---|---|---|
| **Inner Lumen** | 96.8 ± 0.2 % | 94.4 ± 0.4 % |
| **Entire AAA** | 94.8 ± 0.5 % | 89.5 ± 0.6 % |
| **Outer Wall + ILT Only** | 88.2 ± 0.8 % | 85.2 ± 0.9 % |

Model outputs for two patients within the test set are shown in Figure 6 along with their respective ground truths and DICE similarity scores. Major areas of discrepancy arise at the iliac bifurcation and at other points of high curvature (Fig 7B,D).

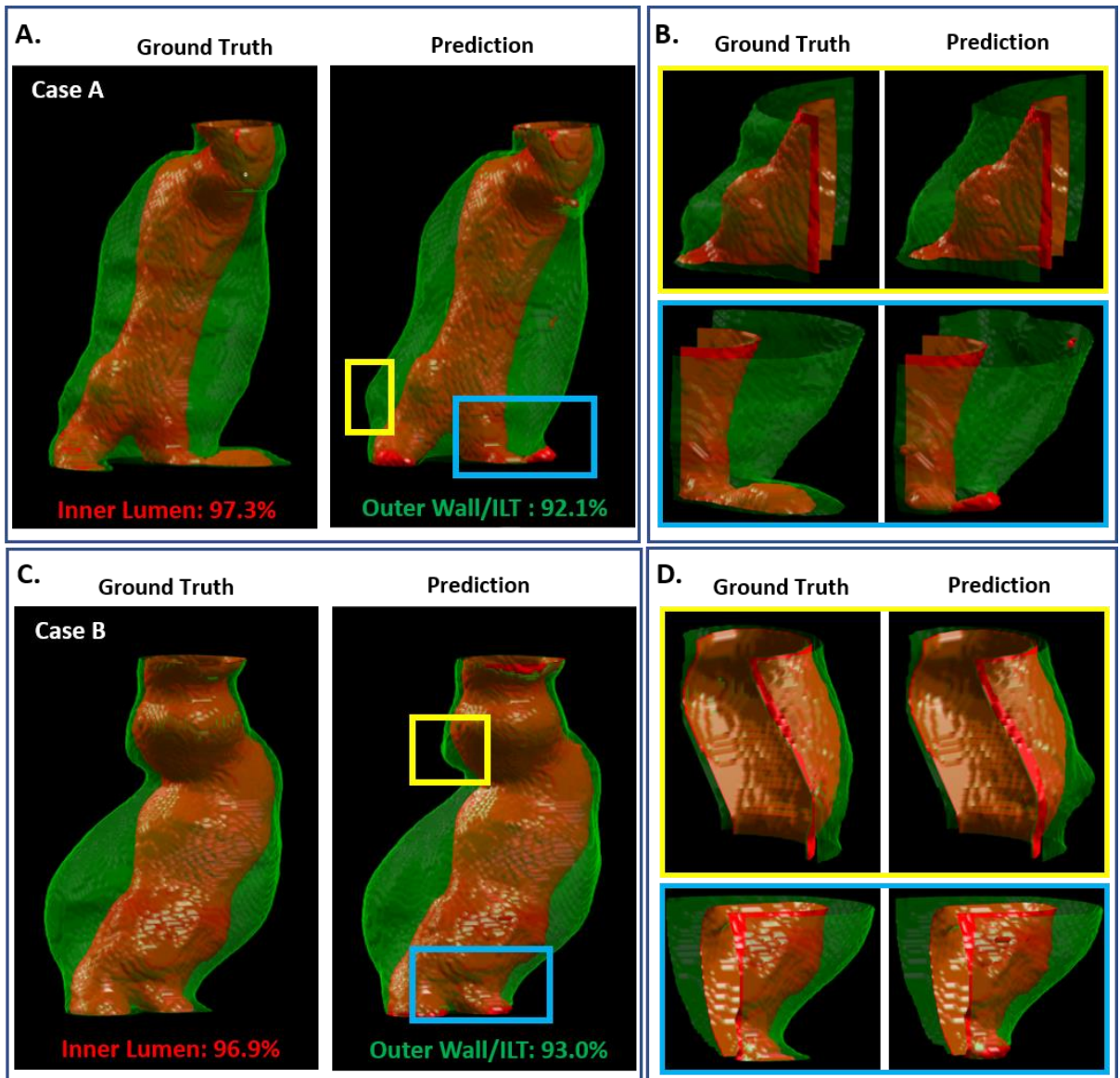

**Figure 7:** Attention-based 3D- U-Net outputs for two patients **(A,C)** with the labelled ground truth masks. DICE scores for both the inner lumen and outer wall predictions are indicated for each patient. Various points of discrepancy within the two predictions are highlighted **(B,D)**.

**Aortic Segmentation from CTA Images**

Figure 8 A-C illustrates the evolving DICE score metric for the validation group during training of *Attn U-Nets A-c* involved in the segmentation of the aorta from contrast-enhanced CTA images. Table 6 displays the performance of *Attn U-Nets A-C* on the ability to segment CTA images within the testing cohort via the DICE score metric. Merging the outputs of *Attn U-Nets B* and *D to* generate the entire aortic volume prediction results in an overall DICE Score accuracy of 93.0 ± 0.6% (Inner Lumen: 96.4 ± 0.3%, Outer Wall: 87.3 ± 0.9%). Figure 8 shows the model predictions for *Attn U-Nets A-C* for a patient within the external test cohort compared against their respective ground truth annotations.

## Aortic Segmentation from Non-Contrast CT Images

Figure 8 D,E illustrates the evolving DICE score metric for the validation cohort during training of *Attn U-Nets D,E,* which are involved in the segmentation of the aorta from non-contrast CT images. Table 7 displays the performance of *Attn U-Nets D-E* on the ability to segment non-contrast CT images within the testing cohort via the DICE score metric. An example segmentation is illustrated in Figure 10.

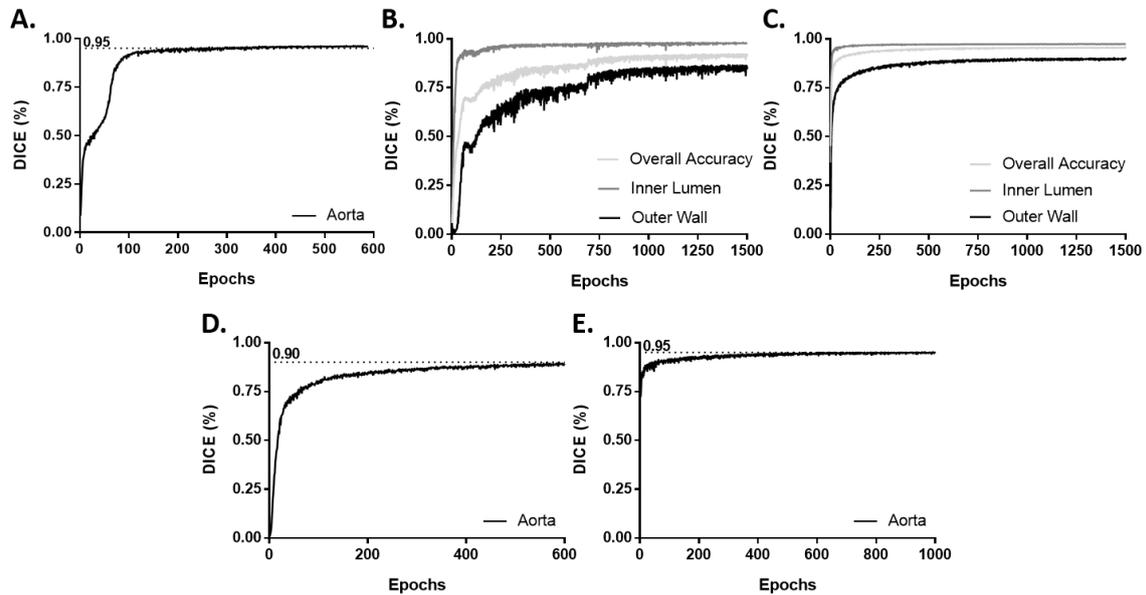

**Figure 8:** Training paradigm for Aortic Segmentation from Contrast/ Non-Contrast CT Images. **A.** *Attn U-Net A* was trained for 600 epochs on down-sampled isotropic CT images. **B-C**. *Attn U-Nets B and C* were trained for 1500 epochs on the ROIs derived from *Attn U-Net A* – aortic arch **(B)**, and descending aorta/AAA **(C)**. **D.** *Attn U-Net D* was trained for 600 epochs on down-sampled isotropic non-contrast CT images. **E.** *Attn U-Net E* was trained for 1000 epochs on ROIs derived from *Attn U-Net D*.

**Table 6:** Aortic Segmentation assessment of Attention-based U-Nets A-C from contrast CT Images.

|  |  | U-Net A | U-Net B | U-Net C |
|---|---|---|---|---|
|  | **Input Data** | Low Resolution, Isotropic CTA | High Resolution, Isotropic CTA Region (Aortic Arch) | High Resolution, Isotropic CTA Region (Desc. Aorta/ AAA) |
|  | **Output Segmentation** | Entire Aorta | Aortic Arch (Lumen, Outer Wall) | Desc. Aorta/AAA (Lumen, Outer Wall) |
| **DICE(%)** | **Combined** | 93.4 ± 1.2 % | 95.8 ± 0.6 % | 94.8 ± 0.5 % |
| | **Inner Lumen** | - | 96.3 ± 0.4 % | 96.8 ± 0.2 % |
| | **ILT/Outer Wall** | - | 84.3 ± 1.2 % | 89.3 ± 0.5 % |

**Table 7:** Aortic Segmentation assessment of Attention-based U-Nets D,E from non-contrast CT images.

|  |  | U-Net D | U-Net E |
|---|---|---|---|
|  | Input Data | Low Resolution, Isotropic Non-Contrast CT | High Resolution, Isotropic Non-Contrast CT (Desc. Aorta/ AAA) |
|  | Output Segmentation | Entire Aorta | Desc. Aorta/AAA (Lumen, Outer Wall) |
| DICE(%) | Combined | 88.7 ± 0.5 % | 93.2 ± 0.7 % |

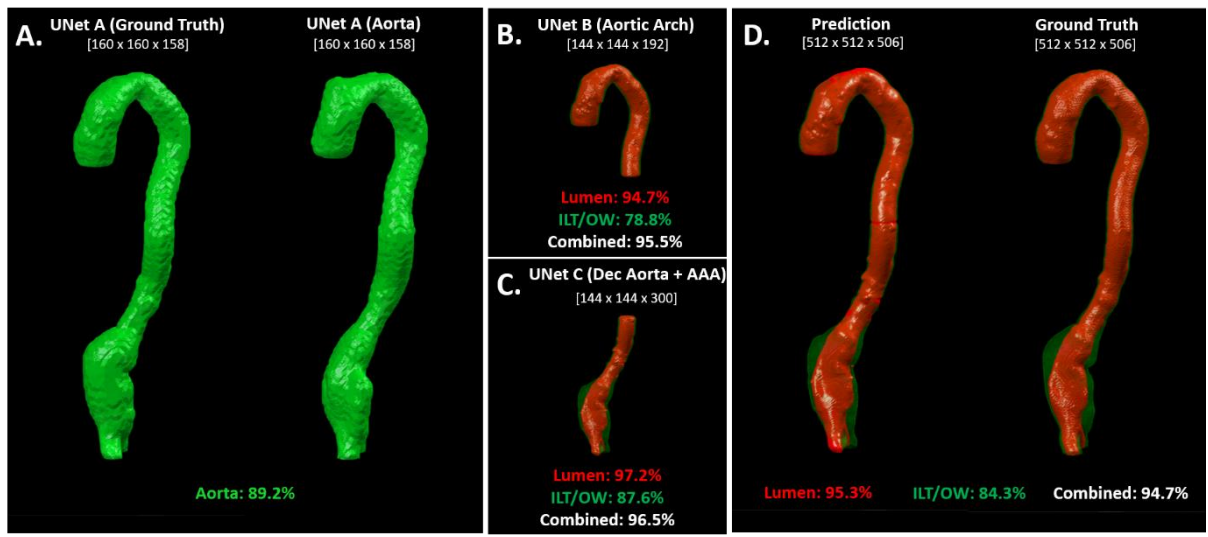

**Figure 9:** Automated aortic segmentation pipeline (*Attn U-Nets A – C*) of a contrast CT image for a patient within the testing cohort. **A.** *Attn U-Net A* identified the aortic structure from down-sampled images and was the basis for ROI detection. **B-C.** *Attn U-Nets B + C* identified both the inner lumen and outer wall predictions for their indicated region. **D.** Region predictions were combined and compared against the overall ground truth to assess overall accuracy.

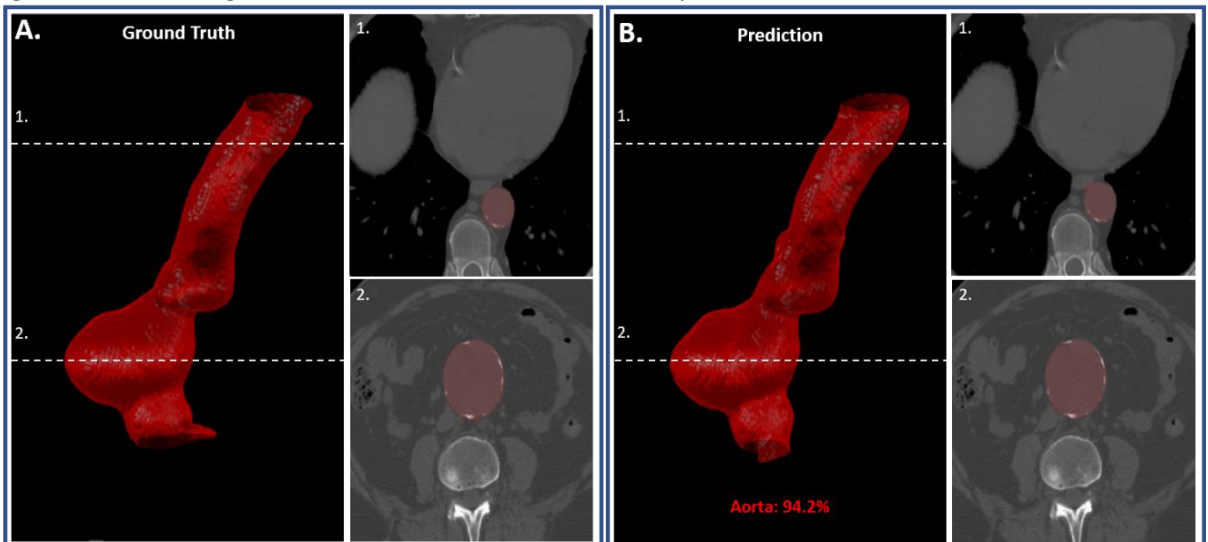

**Figure 10:** Automated aortic segmentation pipeline (*Attn U-Nets D-E*) of a non-contrast CT image for a patient within the testing cohort. **A.** Manually Segmented Aorta as Ground Truth. **B.** Aortic Segmentation as predicted by *Attn U-Net E*. Two regions within the (1.) thoracic region and (2.) aneurysm have been highlighted in both images

**Discussion**

This method employs, for the first time, a fully automatic and high-resolution algorithm that is able to extract the aortic volume from both CTA and non-contrast CT images at a level superior to that of other currently published methods. This extracted volume can be used to standardize current methods of aneurysmal disease management and sets the foundation for subsequent complex geometric analysis. Furthermore, the proposed pipeline can be extended to other vascular pathologies.

Prior to the advent of machine-learning approaches, AAA segmentations were performed using intensity-based semi-automatic algorithms (level-sets, active shape models and graph cut methods) [10-14]. The primary drawback of these methods was the failure to accurately detect the outer boundary of the outer wall/aneurysm as the intensity of this region may be similar to that of adjacent structures. The addition of shape-priors to these models attempts to limit these variations and improve overall prediction accuracy [15]. Although the output of these models may, at times, provide good results, there are significant limitations that prevent them being used within the clinical setting. Most of these methods are semi-automatic and require significant model optimization. Furthermore, these methods require complex user-input, including prior inner lumen segmentation along with centreline extraction, to assist the aneurysmal segmentation algorithm [10, 11, 15]. Finally, many of these early models are highly data-set dependent, which decreases model robustness and the generalizability required for implementation within the clinical environment.

Recently, a variety of machine/ deep-learning methods on contrast-enhanced CTAs have been proposed to tackle this problem without encountering many of the limitations of their predecessors. Variations on Deep Belief and U-net based networks have been used to segment the infra-renal region of the aorta [16, 17]. Unfortunately, many of these networks are limited to 2-D inputs (axial CT slices), which may fail to appropriately capture the 3D geometry of the aneurysm. The accuracy and reproducibility of these models is like that of earlier methods as they are trained and validated on small data sets. Lopez-Linares et al. recently proposed a Holistically-Nested Edge Detection (HED) network trained in both 2D and 3D that out-performs currently available methods in both pre- and post- operative AAA segmentation [18]. However, this method is limited to single-class segmentation of the aneurysmal outer wall and performs poorly with small aneurysms and those with a small thrombus burden.

In the first experiment, we compared the performances between a 3-D U-Net with and without attention gates (*Attn U-Net*) for the segmentation of the aneurysmal region. Current convolutional neural network architectures, as seen in the *U-Net*, are able to capture semantic contextual information by generating a coarse feature-map grid through iterative down-sampling of the input. This way features on this coarse map model location and relationship between structures/tissues at the organ level. However, this method struggles to capture small target objects with increased shape variability. This is especially important for pathological vascular cases.

Integrating attention gates, which is commonly used in natural image analysis and image classification tasks, into the CNN architecture has shown promise in focusing on target structures without the need for additional training/supervision[6]. The primary benefit of this gating mechanism is that it can suppress predictions in irrelevant background regions without requiring the training of multiple models and many model parameters. Additionally, these parameters can be trained simultaneously with the underlying network architecture using standard back-propagation methods.

The strength of this attention-based U-Net has been previously documented on the segmentation of abdominal structures[6]; however, its role in aortic/aneurysmal segmentation has never before been evaluated. The superior segmentation performance with the use of attention gates, as seen in this study, has rationalized its incorporation within the full aortic segmentation pipeline.

Furthermore, we show for the first time, the ability to use a deep learning method to isolate the aorta from a non-contrast CT scan. This will allow for the extraction of complex morphological information from non-contrast images and subsequent longitudinal analysis. The same methodology underpinning this work can be extended to enable automatic segmentation of other hollow or solid organs (such as the kidneys, veins, liver, spleen, bladder, or bowel) with or without the use of intravenous contrast agents.

Although computed tomography angiography may provide unique insight into aneurysm morphology and the structure of the vascular tree, it is not without its disadvantages. Administration of contrast agents for CTAs requires needle insertion. This can cause additional patient discomfort and has been associated with multiple complications including inadvertent arterial puncture by needle, and contrast leak from veins causing skin irritation/damage. Additionally, contrast agents are nephrotoxic and have a 10% incidence of acute kidney injury (contrast-induced nephropathy) following use. This is especially a problem within the elderly population, who either have decreasing baseline renal function or concomitant chronic kidney disease. The use of contrast within this cohort of patients is contraindicated as there is a recognized risk of complete kidney failure. Given that a large sub-cohort of patients with aortic aneurysmal disease may have diagnosed renal disease, this study highlights the necessity to re-evaluate the role of non-contrast CT imaging for the management of aneurysmal disease.

**Conclusion**
We developed a novel automated pipeline to enable high resolution reconstruction of blood vessels using deep learning approaches. This pipeline enables automatic extraction of morphologic features of blood vessels and can be applied for research and potentially for clinical use.

**Acknowledgement**
We acknowledge the support from the following: Medical Sciences Division, University of Oxford Medical Research Fund; John Fell Fund, University of Oxford; Academy of Medical Sciences Starter Grant to RL (AMS_SGL013\1015); Oxford University Claredon Scholarship to AC. We acknowledge the input by Luke Markham and Hao Xu in the preparation phase of this project. The methods described in this manuscript is subject to a patent filing (UK priority filing, P286498GB).